\documentclass[conference]{IEEEtran}
\usepackage{graphicx}
\usepackage{balance}

\makeatletter
\newcommand{\linebreakand}{%
  \end{@IEEEauthorhalign}
  \hfill\mbox{}\par
  \mbox{}\hfill\begin{@IEEEauthorhalign}
}
\makeatother

\begin{document}

\title{Student Mastery or AI Deception? Analyzing ChatGPT's Assessment Proficiency and Evaluating Detection Strategies}

\author{\IEEEauthorblockN{Kevin Wang}
\IEEEauthorblockA{Department of Computer Science\\
University of British Columbia\\
Kelowna, BC, Canada, V1V 2Z3\\
wskksw@mail.ubc.ca}
\and
\IEEEauthorblockN{Seth Akins}
\IEEEauthorblockA{Department of Computer Science\\
University of British Columbia\\
Kelowna, BC, Canada, V1V 2Z3\\
seth.akins@ubc.ca}
\and
\IEEEauthorblockN{Abdallah Mohammed}
\IEEEauthorblockA{Department of Computer Science\\
University of British Columbia\\
Kelowna, BC, Canada, V1V 2Z3\\
abdallah.mohamed@ubc.ca}
\linebreakand
\IEEEauthorblockN{Ramon Lawrence}
\IEEEauthorblockA{Department of Computer Science\\
University of British Columbia\\
Kelowna, BC, Canada, V1V 2Z3\\
ramon.lawrence@ubc.ca}
}

\maketitle

\begin{abstract}

Generative AI systems such as ChatGPT have a disruptive effect on learning and assessment. Computer science requires practice to develop skills in problem solving and programming that are traditionally developed using assignments. Generative AI has the capability of completing these assignments for students with high accuracy, which dramatically increases the potential for academic integrity issues and students not achieving desired learning outcomes. This work investigates the performance of ChatGPT by evaluating it across three courses (CS1,CS2,databases). ChatGPT completes almost all introductory assessments perfectly. Existing detection methods, such as MOSS and JPlag (based on similarity metrics) and GPTzero (AI detection), have mixed success in identifying AI solutions. Evaluating instructors and teaching assistants using heuristics to distinguish between student and AI code shows that their detection is not sufficiently accurate. These observations emphasize the need for adapting assessments and improved detection methods.

\end{abstract}


\section{Introduction}

Since the introduction of ChatGPT, there has been significant research on how generative AI can impact education both positively \cite{baidoo2023education,bullshitChatGPT} and negatively \cite{bullshitChatGPT,LoRapidReview, bordt2023chatgpt,malinkaDegree}. There are concerns related to academic dishonesty with students using AI to complete their assessments, especially in writing courses \cite{bullshitChatGPT}. Generative AI also has problems with fake, inaccurate, and biased data \cite{bullshitChatGPT,LoRapidReview}. 

Assessments in computer science are also under threat. Programming assignments used to improve student problem solving are easily completed by generative AI. Prior work has demonstrated that assignments in introductory CS courses \cite{geng2023chatgpt} are completed with high accuracy using ChatGPT. Further, software development companies have integrated generative AI into integrated code development environments with products such as GitHub Copilot \cite{copilot} and Codeium \cite{codeium}.

In response, educators are developing recommendations to address the issue. These suggestions primarily focus on employing AI models to detect plagiarism, implementing consistent rules and guidelines \cite{loweffort}, and exploring innovative assessment methods \cite{LoRapidReview,variability}. Commonly used plagiarism detection software such as MOSS \cite{mosspaper,MOSS}, JPlag \cite{jplag}, and others \cite{agrawal2016state} are effective at determining similarity matching with other code submissions. However, generative AI creates new code during every prompt request, and it is uncertain whether these tools are still effective. Generative AI detection software such as GPTZero \cite{GPTZeroTechnology} and ZeroGPT \cite{ZeroGPTTechnology} are primarily focused on detecting AI generated writing and have not been extensively evaluated for detecting AI generated code.

In this paper, the effectiveness of ChatGPT in completing assessments in three courses (CS1, CS2, databases) is evaluated. CS1 and CS2 have been explored in other prior work \cite{geng2023chatgpt,apex}, but evaluation in an upper-year database course is new. A novel contribution is examining the effectiveness of current detection software against ChatGPT generated code. The performance of instructors and teaching assistants is also measured, which has not been previously examined. 

Evaluating the effectiveness of human and automated detection of AI generated code reveals challenges in current detection systems and methods. However, there may be opportunities to improve detection by utilizing multiple approaches and some common distinguishing features of AI generated code discovered during the experiment.

This work examines the questions:

\begin{itemize}

    \item What is the performance of ChatGPT in multiple CS courses including CS1, CS2, and databases?
    
    \item What is the effectiveness of existing plagarism and AI detection software for detecting AI generated code?
    
    \item How accurate are instructors and teaching assistants at distinguishing between code created by students and AI?
    
\end{itemize}

\section{Background}

ChatGPT \cite{chatGPT} has demonstrated its effectiveness in assisting programmers by providing debugging support \cite{NigarSolveBugs} and aiding in software development. It introduces novel ideas and builds upon previous AI tools like GitHub Copilot \cite{copilot}, enhancing their capabilities. However, when applied in education, educators and researchers have voiced concerns regarding its use \cite{bullshitChatGPT,LoRapidReview,bordt2023chatgpt,malinkaDegree,geng2023chatgpt} and ability to enable student academic dishonesty.

Generative AI systems like ChatGPT are continually evolving and improving, and various studies have been performed to determine their effectiveness in completing computer science assessments. One result argued that  ChatGPT 4 exhibited minimal understanding of computer science compared to human beings \cite{bordt2023chatgpt}. Many other studies have demonstrated much more positive results. Geng {\em et al.} tested ChatGPT in an introductory functional course and discovered that its performance ranked in the middle of the class, achieving a B- grade \cite{geng2023chatgpt}. Malinka {\em et al.} tested ChatGPT and demonstrated how easy it is to cheat in IT security related education \cite{malinkaDegree}. ChatGPT has excellent performance in introductory CS courses \cite{apex} and excels in many standardized exams \cite{openai2023gpt4}.

How generative AI should be integrated into computer science courses is a challenging discussion. CS graduates will require the use of these tools in industry, and tools such as GitHub Copilot can save significant developer time and increase productivity. However, using ChatGPT and other generative AI systems on education assessments eliminates any student learning and achieving learning outcomes. The AI is completing the assessment rather than the student learning. Instructors require clear communication on acceptable generative AI use in their courses. Limiting academic dishonesty is possible by assignment modifications \cite{variability} and open communication with students on integrity \cite{loweffort}.

Academic misconduct is common with research studies finding that 23\% to 89\% of students admit to cheating \cite{miller2011reasons}. In \cite{Bazaluk2023}, 92.7\% of students would copy the full assignment from someone when asked how they would cheat. Students cite the potential of getting caught ranks high among the reasons not to cheat \cite{miller2011reasons}. As a result, the detection of plagiarism still plays an important role in keeping students from cheating. 

A literature review \cite{aniceto2021source} evaluated methods to detect source code plagiarism, such as MOSS \cite{mosspaper,MOSS}, JPLAG \cite{jplag}, and SIM \cite{mosspaper,agrawal2016state}. Most methods are based on different algorithms to detect similarities between codes.  MOSS uses a technique called winnowing, which fingerprints code files to detect whether other files are similar. Source code detection has also been attempted with machine learning techniques to increase the efficiency and effectiveness of detection \cite{agrawal2016state,engels2007plagiarism}. The goal of these tools are to check the level of similarity of programs. 

One response to cheating with AI is to use detection software to determine AI generated assessments. Plagiarism detection software for code has not been extensively evaluated. GPTZero is one example among the numerous detection tools currently available \cite{GPTZeroTechnology}. GPTZero's detection is based on burstiness and perplexity. Burstiness refers to the unusual frequency of certain words or phrases in a short span of text, which can indicate artificial generation or manipulation. Perplexity measures how well a probability distribution predicts a sample \cite{GPTZeroTechnology}. It is hard to distinguish between human and AI generated text \cite{bullshitChatGPT,LoRapidReview}. An evaluation of generative AI detection tools for written computer science reports \cite{orenstrakh2023detecting} indicates various levels of detection accuracy and many false positives, with very poor performance on code submissions. 

The ability for a commercial software, APEX, to detect ChatGPT answers was evaluated \cite{apex} for CS1. The results were that ChatGPT had excellent performance, and ChatGPT solutions could be detected by a variety of methods such as style difference, limited time spent, and similarity checking if a sufficient number of students submitted answers created by ChatGPT. There is a need to evaluate free-to-use tools such as MOSS \cite{MOSS} and JPlag \cite{jplag} to quantify their detection accuracy.

\section{Methodology}

\subsection{Courses Overview}

The effectiveness of generative AI, specifically ChatGPT 3.5, was evaluated for three common CS courses: CS1, CS2, and databases (DB) at a large, research-intensive university.

{\bf CS1} covers the basics of programming including problem-solving, algorithm design, data and procedural abstraction, and the development of working programs in the Java language. Concepts include variables, operators, expressions, control statements (if, switch, loops), arrays, methods, object-oriented programming, and debugging. The course includes 9 programming assignments. 

{\bf CS2} covers topics such as object-oriented programming, abstract classes and interfaces, exception handling, input/output streams, recursive methods, data structures, Java generics, and sorting algorithms. The course includes 9 programming assignments. 

{\bf Databases (DB)} is an introduction to databases with topics including querying with SQL, designing databases using ER and UML modeling, and programming with Java/JDBC. The course has 10 assignments and two computer-based midterm exams.

\subsection{Evaluating ChatGPT Performance}

The experiment focuses on the capabilities of students using ChatGPT and whether they can achieve full marks without prior knowledge of the related materials. We used the assignment and exam questions as prompts for ChatGPT to generate answers. ChatGPT was provided with all related information that is normally given to students to start the assignments, for example, a sample run of the solution. The generated answers were then compared against the official solutions to assign grades. The performance on DB exams was also evaluated as they were done online while  CS1 and CS2 were paper exams. We copied exam questions with the given context (for example, DDL for a set of SQL questions) and graded the responses. Given the objective of assessing students' ability to obtain assignment answers without possessing course related knowledge, follow-up non-technical questions were permitted for ChatGPT if the initial answers were unsatisfactory.

\subsection{Evaluating AI Detection}

The effectiveness of detecting AI generated code was performed using two experiments. The data set for both experiments consisted of 50 anonymized student submissions for CS1 and CS2 assignments and 50 different AI generated answers for those assignments. The student submissions were from a previous course offering. All identifying information was extracted from the submissions, and students were not involved in the research in any way as the course was completed in the past. The different AI generated solutions were collected by providing the given starter code, questions, sample runs, and rubrics to ChatGPT in different conversations. 

The automated AI detection experiment evaluated two types of detection tools on this data set. The similarity tools included MOSS \cite{mosspaper,MOSS} and JPlag \cite{jplag} designed specifically for detecting code plagiarism, while the AI detection tool, ChatGPT Zero, aims to identify AI generated content. For MOSS and JPlag, we provided a randomized folder of mixed assignments, analyzed the returned similarity scores, and used a similarity cutoff to determine if a submission was human or AI. For GPTZero, we provided 24 solutions (14 human, 10 AI) and asked GPTZero to determine the source. 

The second experiment measured human capacity for detecting AI generated code. Course teaching assistants and instructors attempted to distinguish between anonymous student solutions and those generated by ChatGPT in a randomized selection  of AI generated and student work. The randomized selection includes 32 solutions from 4 assignments in CS1 and CS2 with 14 solutions generated by AI. This information was collected via survey. There was a total of 10 survey responses.

\section{Results}

\subsection{ChatGPT Performance}

ChatGPT is very effective at solving assignments in CS1 and CS2 (see Tables \ref{tab:cs1} and \ref{tab:cs2}). It correctly outputs the correct code for every coding question in CS1 on the very first try. Each assignment may have multiple distinct questions as shown in the {\bf $\#$Q} column. The only question it could not do is a small part in A6 where students are asked to submit screenshots of the debugging process. Assignments for CS2 are also completed with near 100\% accuracy. Close to zero coding skills are required to get the correct answers for any of these assignments.  However, certain questions that are longer and with extensive starter code require students to feed information to ChatGPT in multiple cells. The mark for CS2 A1 was lower as one question asked to display the inheritance graph, which required more explanation of the inheritance relationships for ChatGPT to provide correct answers. 

\begin{table}[ht]
\centering
\begin{tabular}{|c|c|l|c|}
\hline
 & {\bf \#Q} & {\bf Topic} & {\bf Grade}   \\
\hline
A1 & 3& Writing your first Java programs & 100\% \\
A2 & 4& Variables \& data types & 100\% \\
A3 & 5& Arithmetic Operators& 100\% \\
A4 & 5& Relational and logical operators,& 100\% \\
A5 & 4& Loops (while, do-while, for) & 100\% \\
A6 & 4& Methods \& Debugging & 83\% \\
A7 & 3& Arrays \& passing arrays to methods & 100\% \\
A8 & 3& Multi-dimensional arrays & 100\% \\
A9 & 2& OOP & 100\% \\
\hline
\end{tabular}

\caption{ChatGPT Performance in CS1}
\label{tab:cs1}
\end{table}

\begin{table}[ht]
\centering
\begin{tabular}{|c|c|l|c|}
\hline
 & {\bf \#Q} & {\bf Topic} & {\bf Grade}   \\
\hline
A1 & 2& OOP & 96\% \\
A2 & 4& Abstract Classes and Interfaces & 100\%\\
A3 & 4& Exception Handling \& I/0 & 100\%\\
A4 & 3& Binary I/O & 100\%\\
A5 & 4& Recursion & 100\%\\
A6 & 2 & ArrayLists & 100\%\\
A7 & 1& Java Collection Framework & 100\%\\
A8 & 1& Java Collections (LinkedLists) & 100\%\\
A9 & 1 & Sorting & 100\%\\
\hline
\end{tabular}

\caption{ChatGPT Performance in CS2}
\label{tab:cs2}
\end{table}

\begin{figure}[!ht]
	\centering
	\includegraphics[width=3.3in]{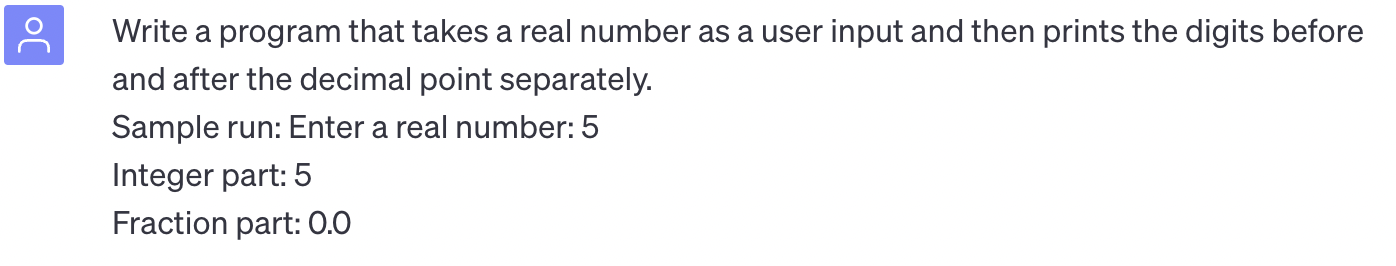}
        \caption{Prompt for CS2 A3 Question 2}
        \label{fig:prompt}
\end{figure}

\begin{figure}[!ht]
	\centering
	\includegraphics[width=3.3in]{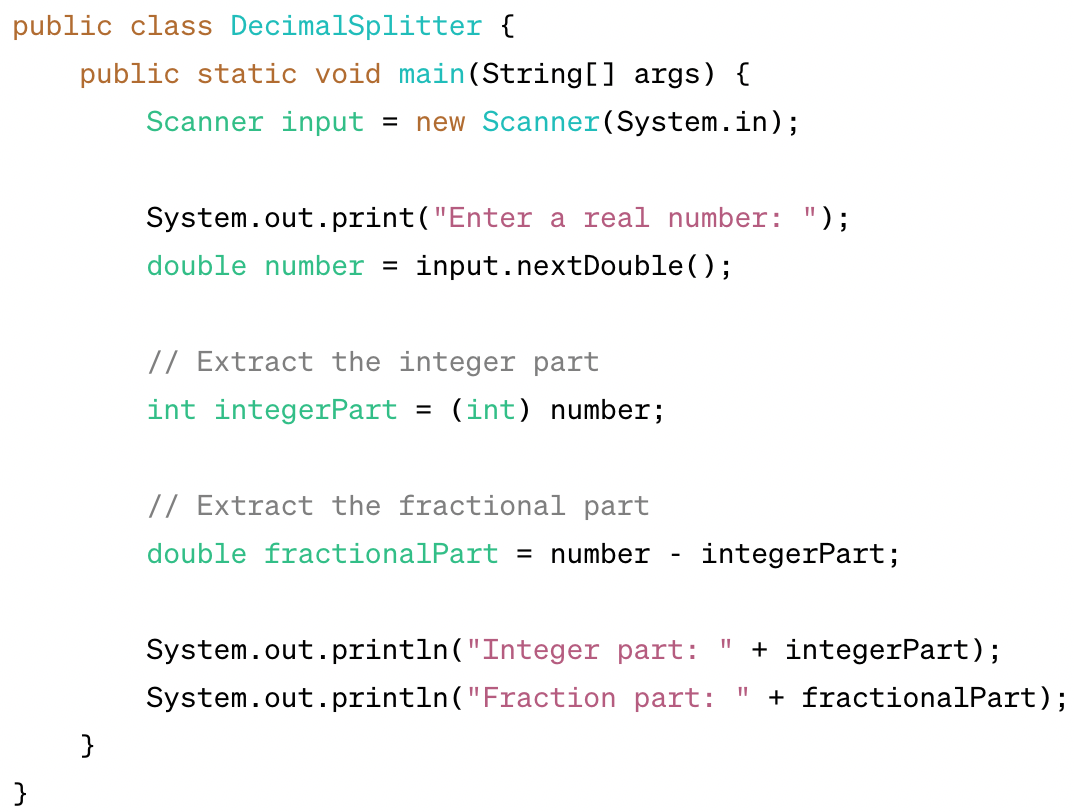}
        \caption{ChatGPT Response}
        \label{fig:response}
\end{figure}

A sample prompt is in Figure \ref{fig:prompt}, and the answer in Figure \ref{fig:response}. The returned response is correct and even includes comments. The performance is expected as questions in CS1 and CS2 must be very clear for students to understand and often contain example output. This clarity makes excellent prompts for ChatGPT and requires only that a student copy the question directly as their prompt.

The database (DB) course is an upper-year course with a wider range of materials and questions. ChatGPT has very high performance with excellent marks (see Table \ref{tbl:db}) for labs 1, 2, 3, and 9 on relational algebra, SQL, and XML/JSON. The code based labs of 6, 7, 8, and 10 require completing hundreds of lines of code across multiple files. Upon providing the template code, DDL, and the rubrics, ChatGPT gets extremely close to providing the exact answers, often only pending small changes such as database connections. 

\begin{table}[ht]
\centering
\begin{tabular}{|c|l|c|}
\hline
 & {\bf Topic} & {\bf Grade}   \\
\hline
A1 & Relational Algebra& 100\% \\
A2 & Creating tables & 100\% \\
A3 & Writing SQL queries& 100\% \\
A4 & UML Modeling & 0\%\\
A5 & Converting UML Diagrams& 0\%\\
A6 & Database and programming & 100\% \\
A7 & Building a Database-enabled Web Site& 85\% \\
A8 & Images, Security, and Transactions& 100\% \\
A9 & XML, JSON, Views, and Triggers& 100\% \\
A10 & Final project with database & 90\%  \\
midterm1 &28.5/30 & 95\% \\
midterm2 & 17/30  &  57\% \\
\hline
\end{tabular}
\caption{ChatGPT Performance in DB}
\label{tbl:db}
\end{table}

\begin{figure}[!ht]
	\centering
	\includegraphics[width=3.3in]{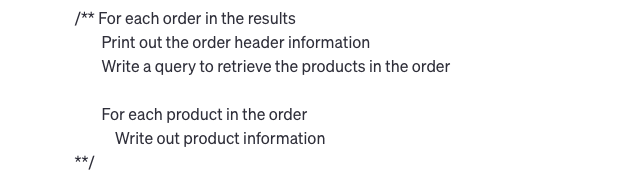}
        \caption{ChatGPT prompt for DB Lab7 Q2}
        \label{fig:304prompt}
\end{figure}

As an example, a snippet of the prompt in Figure \ref{fig:304prompt} is copied from the starter code provided to students. The prompt is accompanied with the database schema and grading rubric for the assignment. 

\begin{figure}[!ht]
	\centering
	\includegraphics[width=3.3in]{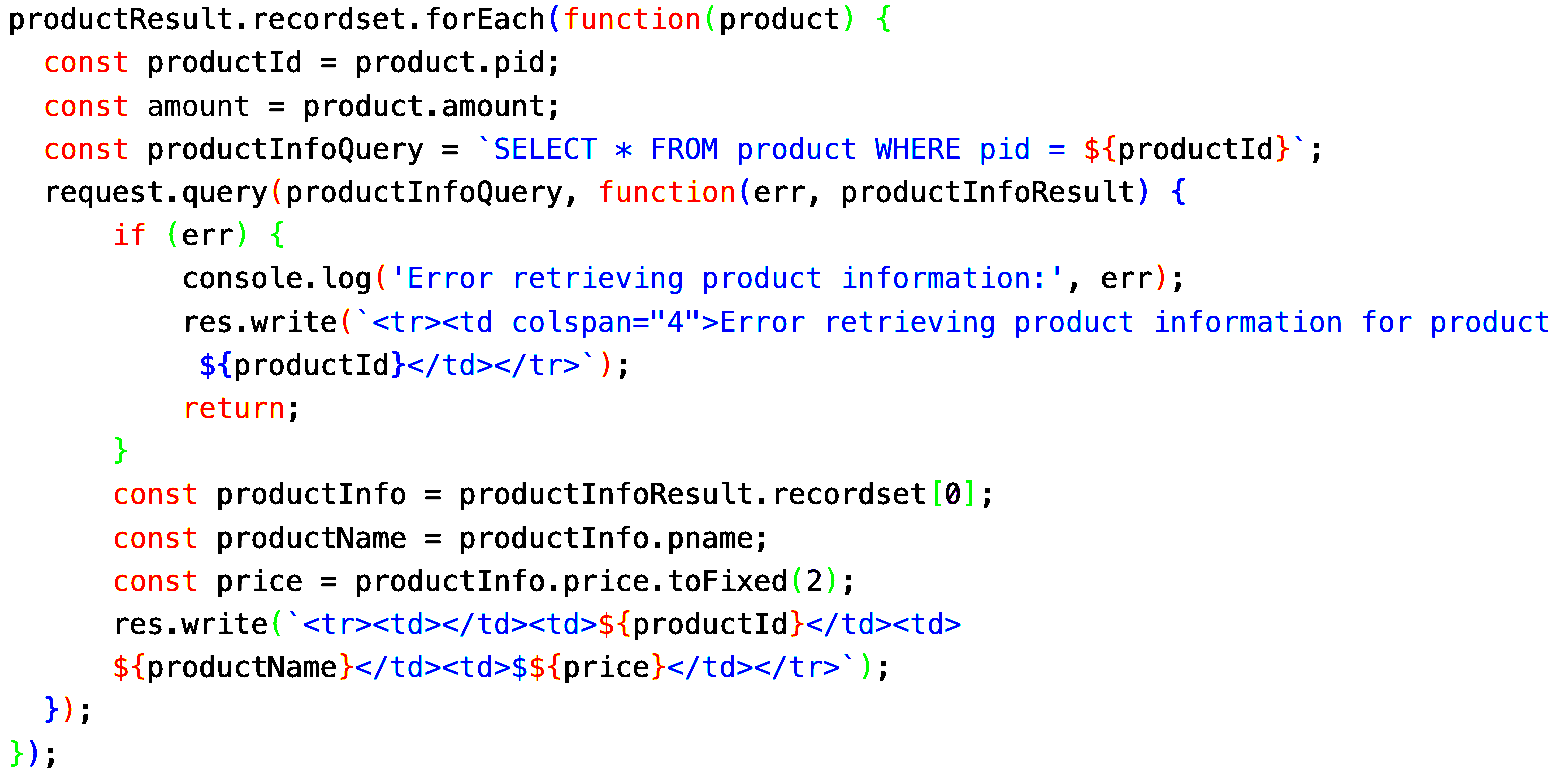}
        \caption{ChatGPT answer for DB, Lab7, Q2}
        \label{fig:304response}
\end{figure}

ChatGPT's response in Figure \ref{fig:304response} is correct except that it queried from the product table instead of order summary data. The structure of the code is correct. It is also noteworthy to include that upon asking a follow up question, ``from ordersummary", ChatGPT returned the correct code solution.

\begin{figure}[!ht]
	\centering
	\includegraphics[width=2in]{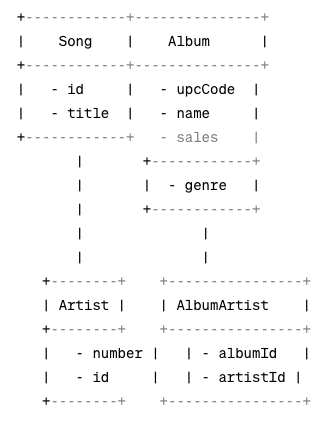}	
        \caption{ChatGPT answer for ER/UML Diagram}
        \label{ER}
\end{figure}

However, ChatGPT does not currently draw ER/UML diagrams as well as required in labs 4 and 5. It is still able to come up with some basic diagram (see Figure \ref{ER}). Utilizing unique questions in UML such as in \cite{ramon2022} allows for auto-grading while being resistant to generative AI.

ChatGPT tested extremely well in midterm 1, obtaining a 28.5/30, while getting only 17/30 on midterm 2 due to the 10 points allocated to ER diagrams.

\subsection{Automated AI Detection}

\subsubsection{Similarity}

We conducted a series of tests to evaluate the effectiveness of similarity detection systems MOSS and JPlag. For four different questions in CS1 and CS2, 100 submissions were given to the systems. The percentage of AI submissions varied from 5\% to 50\%. A submission was flagged as AI if its highest match was with an AI solution, and it was above a similarity cutoff (90\% for JPlag, 70\% for MOSS).

\begin{figure}[!ht]
    \centering
    \includegraphics[width=3.3in]{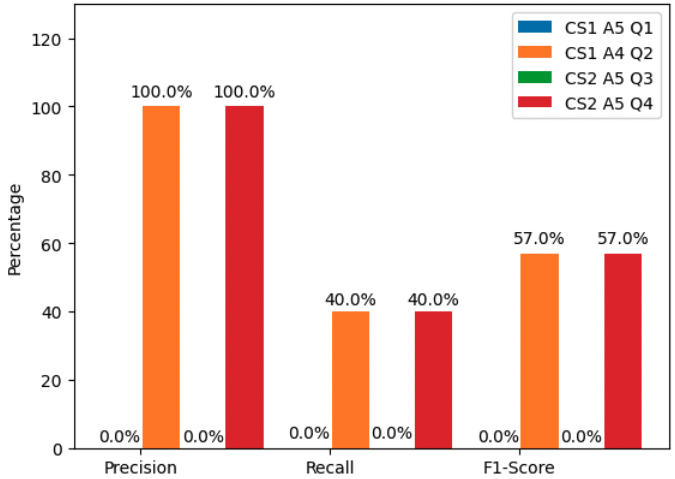}
    \caption{JPlag Performance for 5\% AI Submissions}
    \label{fig:5perc}
\end{figure}

\begin{figure}[!ht]
    \centering
    \includegraphics[width=3.3in]{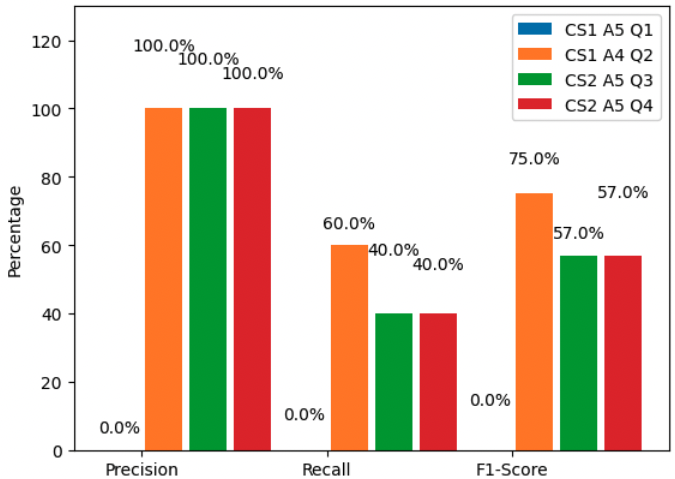}
    \caption{MOSS Performance for 5\% AI Submissions}
    \label{fig:5perc2}
\end{figure}

The precision, recall, and F1 scores for these tests are shown. Figures \ref{fig:5perc} and \ref{fig:5perc2} contain data when 5\% of the solutions are AI for JPlag and MOSS. With a small number of submissions, the AI detection is weak. Some AI solutions are detected, but the recall is lower.

\begin{figure}[!ht]
    \centering
    \includegraphics[width=3.3in]{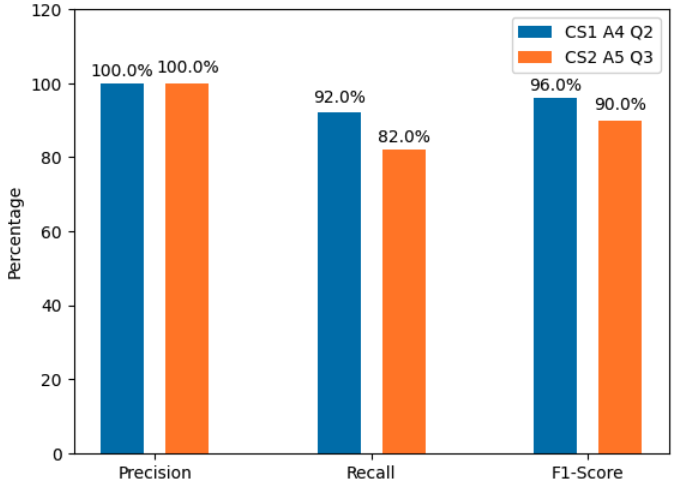}
    \caption{JPlag Performance for 50\% AI Submissions}
    \label{fig:50perc}
\end{figure}

\begin{figure}[!ht]
    \centering
    \includegraphics[width=3.3in]{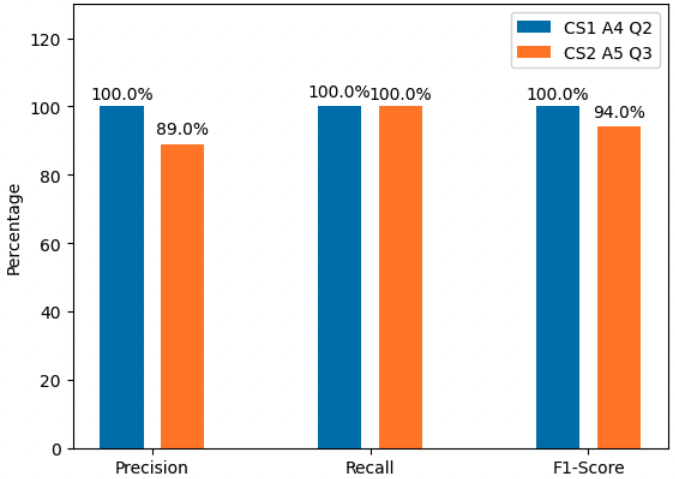}
    \caption{MOSS Performance for 50\% AI Submissions}
    \label{fig:50perc2}
\end{figure}

When the percentage of AI solutions is 50\% (shown in Figures \ref{fig:50perc} and \ref{fig:50perc2}), the detection accuracy is significantly higher. Even though ChatGPT generates different versions on each request, after a sufficient number of requests it is more likely a similar result will be generated.

Several other factors negatively influence the detection process. Students may modify the AI generated solutions and utilize different prompting methods. There are also multiple generative AI systems that can be used. The complexity of the questions is another key factor. Simpler questions lead to more uniform solutions, making it easier to detect AI generated content. More complex questions, with a wider range of responses, make detection more challenging.
   
The number of AI generated submissions is a key factor in their detection, but in a real class environment, it is unknown the number of AI submissions and the generative AI system used. Although an instructor could potentially provide a set of labeled AI submissions for each question, this may not always be practical. However, without this AI sample set, the accuracy of similarity detection systems is below their effectiveness for other forms of assignment copying and plagiarism. The detection of AI generated solutions is not yet reliable enough across different scenarios. Further research is needed to improve AI solution detection using similarity systems.

\subsubsection{AI Detectors}

AI detection tools primarily focused on writing have low success in detecting AI generated code. Providing AI solutions to GPTZero \cite{GPTZeroTechnology} resulted in very low probability of AI scores. The AI generated solutions raised {\em less} suspicion than human solutions, and overall these percentages are way below the threshold to identify AI generated content. The metrics used to detect AI generated writing are not effective when applied to code solutions.

\subsection{Instructor Detection of AI Submissions}

We also evaluated if instructors and teaching assistants are more effective at detecting AI submissions. Course instructors and TAs completed a survey that contained 32 mixed student and AI solutions for 4 assignments across CS1 and CS2. 

\begin{figure}[!ht]
    \centering
    \includegraphics[width=3.3in]{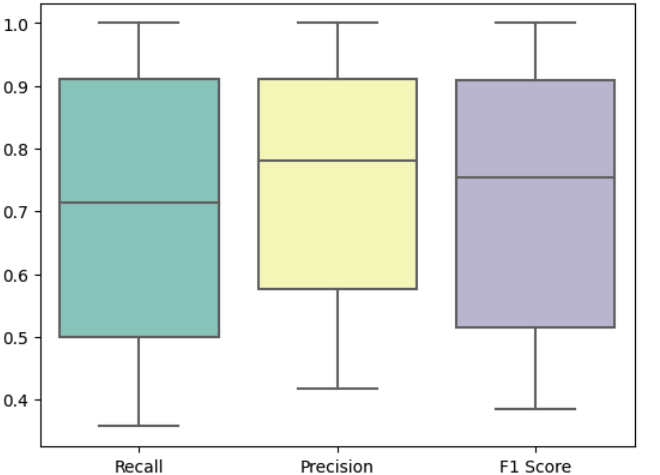}
    \caption{Distribution of Recall, Precision, and F1 Scores for Participants}
    \label{fig:distribution}
\end{figure}

Figure \ref{fig:distribution} contains the distribution of participant performance. The average recall across all participants was 69.29\%. The average precision was 74.51\%, and the average F1 score was 71.42\%.
Overall the detection accuracy rate is 69\%, and the false positive rate is 22\%. 

There are two key observations from these results. First, there is substantial variability between instructors in detecting AI. The lowest F1 score was 40\% with the highest at 100\%. Second, although the performance is better than automated detection software when dealing with a small number of AI submissions, it is still not sufficiently precise for widespread implementation. The average accuracy rate is too low and the false positive rate is too high, which would result in missing many AI submissions, and even worse, falsely accusing students of submitting AI answers.

Participants identified heuristics that had various levels of effectiveness:

\begin{itemize}
    \item Errors in solutions, from coding errors to spelling errors
    \item Scanner not closed
    \item The solution contains methods that are not taught yet in the course or unknown to the instructor
    \item Auto-generated stub from Eclipse IDE
    \item Inconsistent style, spacing, and formatting
\end{itemize}

Common approaches often focus on style aspects of the code as AI generated solutions have a professional style and consistency in grammar and skill-level. Interestingly, it is these style aspects, especially related to variable names and whitespace, that are ignored in similarity matching software. The instructor detection heuristics complement similarity detection systems.

70\% of participants reported that the detection task was very time consuming. As a result, while instructors and TAs can use heuristics to perform detection, it is a labor intensive task that is hard to scale with inconsistent performance depending on the instructor's characteristics. There was a higher performance for instructors and teaching assistants with more experience.

\section{Discussion}

The evaluation of the effectiveness of ChatGPT was done on existing assessments in CS1, CS2, and databases. The questions were copied directly into ChatGPT 3.5 as prompts, and the answers evaluated. Our experiment highlights ChatGPT's impressive capability to generate solutions, even in higher-level courses such as databases. 

In the past, methods of academic dishonesty usually involved copying assignments from peers or online resources. ChatGPT has expanded these methods to include AI generated responses that are no longer copies of previous works, but newly crafted responses. Existing detection methods have mixed success with similarity detection affected by the number of AI submissions and the complexity of the question. Instructor detection using heuristics has variable accuracy depending on instructor experience, and the heuristics may be defeated by minor code format edits in many cases. One useful flag is that ChatGPT lacks an understanding of the course's specific coverage and may use advanced techniques or style formatting that is inconsistent with expected student knowledge.

The overall accuracy of AI detection is lower and less reliable than detecting previous forms of plagiarism. However, there may be opportunities to increase performance by combining multiple methods of similarity detection utilizing pre-generated AI answers for comparison and automatic application of code style heuristics utilized by instructors. The detection problem will continually be evolving and challenging given the rapid improvement of generative AI systems.

\section{Limitations of Research}

ChatGPT's performance depends on the quality of its prompts. Detailed questions result in excellent prompts for ChatGPT. The CS1 and CS2 questions used in this work had a high-level detail that is common in first year courses. More open-ended questions may result in lower performance, or at least, a more conversational effort by the student with ChatGPT to arrive at the correct answer.

The number of AI submissions evaluated for similarity detection was done using multiple conversations with ChatGPT 3.5 with the same prompt. Detection accuracy is expected to be lower if students utilize ChatGPT 4 and other systems as well as varying the prompts provided. Further, no modifications or obfuscations were performed, which would further reduce similarity.

Human detection was done by experienced course teaching assistants and instructors, but the participants had various levels of previous experience with AI and plagiarism detection and were provided with no training.  It would be interesting to see whether providing training and heuristics to use improves detection effectiveness. 

\section{Future Work and Conclusions}

This work evaluated generative AI performance in CS1, CS2, and databases with ChatGPT having a near-perfect score in CS1 and CS2. Even the higher level course on databases had many questions with 100\% performance including larger, team-based coding projects. Assessments with some resistance to ChatGPT were more open-ended, involved images or student-specific outputs, or did not specify expected output in detail.

Given this high performance, students will be tempted to utilize generative AI for completing assessments. Detecting AI submissions has varying accuracy and precision whether done by instructors or similarity detection software. The detection heuristics used by instructors typically focus on style and grammar as AI generated code often appears more professional and better organized. Although these heuristics are moderately useful, they can easily be beaten as students adapt, and more importantly, the false positive rate is too high. Similarity detection software can have good performance if provided with enough AI solutions for comparison to student submissions, but the number of such submissions is affected by problem complexity and diversity of AI responses.

Future work may investigate combining similarity detection with instructor heuristics to determine if higher accuracy is possible. Improved detection may require examining the entire submission history of a student to detect submissions that are potentially beyond their current skill level.

\bibliographystyle{IEEEtran}
\balance
\bibliography{ref.bib}

\end{document}